\documentstyle[12pt, aaspp, epsf]{article}
\catcode`@=11
\def\eqalign#1{\null\,\vcenter{\openup\jot\m@th
  \ialign{\strut\hfil$\displaystyle{##}$&$\displaystyle{{}##}$\hfil
      \crcr#1\crcr}}\,}
\def\eqalignleft#1{\null\,\vcenter{\openup\jot\m@th
  \ialign{\strut$\displaystyle{##}$\hfil&$\displaystyle{{}##}$\hfil
      \crcr#1\crcr}}\,}
\catcode`@=12

\def\lax    {\ifmmode{_<\atop^{\sim}}\else{${_<\atop^{\sim}}$}\fi}
\def\gax    {\ifmmode{_>\atop^{\sim}}\else{${_>\atop^{\sim}}$}\fi}
\def\kms    {\ifmmode{{\rm ~km~s}^{-1}}\else{~km~s$^{-1}$}\fi}

\def\eg{{\it e.\thinspace g.~}}

\def\bk{\lower 6pt\hbox{${\buildrel k\over \sim}$}}
\def\bv{\lower 6pt\hbox{${\buildrel v\over \sim}$}}
\def\ul#1{$\underline{\smash{\vphantom{y}\hbox{#1}}}$}

\def\blankline  {\vskip10truept}

\def\prt {\partial}

\begin{document}

\vfill
\centerline{\bf  QUASI-STATIONARY STATES OF DUST FLOWS}

\centerline{\bf UNDER POYNTING-ROBERTSON DRAG: }

\centerline{\bf NEW ANALYTICAL AND NUMERICAL SOLUTIONS}
\vskip 20pt

\baselineskip 18pt

\author{\bf Nikolai N. Gor'kavyi\altaffilmark{1}}
\affil{Simeiz Department, Crimean Astrophysical Observatory, Simeiz
334242, Ukraine
}
\altaffiltext{1}{e-mail: gorkav@astro.crao.crimea.ua}
\author{\bf Leonid M. Ozernoy\altaffilmark{2}}
\affil{5C3, Computational Sciences  Institute and Department of Physics
\& Astronomy,\\ George Mason U., Fairfax, VA 22030-4444; also Laboratory for
Astronomy and Solar\\ Physics, NASA/Goddard Space Flight Center, Greenbelt,
MD 20771}
\altaffiltext{2}{e-mail: ozernoy@hubble.gmu.edu; ozernoy@stars.gsfc.nasa.gov}
\author{\bf John C. Mather\altaffilmark{3}}
\affil{Code 685, Laboratory for Astronomy and Solar Physics, Goddard Space
Flight Center, Greenbelt, MD 20771}
\altaffiltext{3}{e-mail: mather@stars.gsfc.nasa.gov}
\author{\bf Tanya Taidakova\altaffilmark{4}}
\affil{Simeiz Department, Crimean Astrophysical Observatory, Simeiz
334242, Ukraine
}
\altaffiltext{4}{e-mail: tat@astro.crao.crimea.ua}

\bigskip
\begin{center}
{\it Received 1997 March 21, accepted 1997 May 8}
\end{center}

\newpage
\begin{abstract}
The effect of solar or stellar radiation on dust particles' trajectories (the 
Poynting-Robertson drag) has been studied by a number of authors and applied to
interplanetary dust dynamics in {\it numerical computations}. Meanwhile some
important features of dust flows can be studied analytically by implementing
our novel hydrodynamical approach to use
the continuity equation written in the particle's orbital elements as
coordinates (Gor'kavyi, Ozernoy, \& Mather 1997). By employing this approach
and integrating the continuity equation, we are able to find two integrals
of motion when the Poynting-Robertson drag dominates the dissipative forces 
in the dust flow. These integrals of motion enable us to explore basic
characteristics of dust flows from any sources in the Solar system (such as
asteroids, comets, Kuiper belt, etc.) or in another planetary system. In
particular, we have reproduced the classical solution $n(r)\propto r^{-1}$
that approximately represents the overall distribution of dust in the Solar
system. We have also investigated various factors that could be responsible
for the
deviations of the power law index in $n(r)\propto r^{\delta}$ from $\delta=
-1$, including the influences of the orbital characteristics of dust sources, 
the
evaporation of dust particles, as well as mixtures of dust particles of both
asteroidal and cometary origin. We have calculated the masses and number
densities of asteroidal and cometary components of the zodiacal
cloud at different distances from the Sun.

\end{abstract}

\blankline
{\it Subject headings:} interplanetary medium -- infrared: solar system
\newpage

\section*{\bf 1. INTRODUCTION}
\vskip 6pt
     The Earth orbit is immersed in a multicomponent cloud of interplanetary 
dust and particles of various sizes. The structure and dynamics of
this cloud are very important both for space  missions {\it} and 
interpretations of numerous astronomical data. Small dust particle
scattering (the zodiacal light) and emission contribute substantially to 
infrared data at $\lambda\lambda~ 1-100~\mu$m (Hauser 1995).

The dynamics and
evolution of interplanetary particles are determined by several basic effects
which include: (i) the Poynting-Robertson (P-R) drag; (ii)
resonance effects associated chiefly with Jupiter, Mars, Earth, and
Venus; (iii) gravitational encounters with these planets, which occur
in the form of elastic gravitational scattering of the particles by
the planets; (iv) mutual collisions of the particles,
and (v) evaporation of dust particles.
As we have shown earlier (Gor'kavyi, Ozernoy, \& Mather 1996, 1997),
the dust flow evolution governed by those processes can be conveniently
described by the continuity equation written in the space of orbital 
coordinates:
$$
{{\prt n(x_i,t)\over\prt t}}+ {{\prt\over \prt x_i}}\left (nv_i\right)
=N^+(x_i,t)-N^-(x_i,t), \eqno (1)%
$$
where $i=1...6,$ and $v_i$ is velocity along the $i$-th axis in the phase
space.
In this equation, the {\it div}-term $\prt (nv_i)/\prt x_i$ describes
{\it slow} processes of change of particle orbital elements such as the
particle transport due to both the P-R drag and the resonance
effects. The terms $N^+(x_i,t)$ and $N^-(x_i,t)$ are responsible for {\it %
fast} processes such as the gravitational scattering of particles by the
planets and the contact impacts of particles with each other and with the
planets.

     Let us address the evolution of interplanetary particles
governed by only two processes: (i) the P-R drag that
 {\it continuously} changes the particle's orbit; and (ii) the
gravitational scattering of particles by the planets, which changes the 
particle's orbit like a {\it jump} (for detail, see Gor'kavyi, Ozernoy, 
\& Mather 1997 referred to hereinafter as GOM).
We introduce a small parameter $\epsilon = \delta
n_{sc}/\delta n_{PR} \ll 1$, where $\delta n_{PR}$ and $\delta
n_{sc}$ are the changes, per unit time, due to the appropriate processes
in the number density of dust particles at
 the typical point of the $(a,e)$-space, $a$ being semimajor axis and $e$ being
eccentricity. This small
parameter makes it possible to derive, by the Chapman-Enskog
approach, a set of equations to get the 
first two consecutive approximations for particle density of the flow:
$$
{{\prt n_{PR}(x_i,t)\over\prt t}}+ {{\prt\over\prt x_i}}\left (n_{PR}
v_i\right) =0, \eqno (2)%
$$
$$
{{\prt n_{sc}(x_i,t)\over\prt t}}
=N^+(n_{PR},x_i,t)-N^-(n_{PR},x_i,t). \eqno (3)%
$$
Eq. (2) gives a first approximation to particle density associated with
P-R drag, which is then used to get, with help of Eq. (3), a correcting term
associated with the (second order) process of gravitational scattering.
Eq. (3) applied to the dynamics of particles governed by gravitational
scattering was analyzed in our previous paper (GOM). In the present paper, we 
emphasize
Eq. (2) by exploring the Poynting-Robertson drag as the leading effect
for the dynamics of small particles. The paper is organized as 
follows: In Ch. 1,
we derive the integrals of motion of dust particles, assuming the
PR-drag as the dominating effect. In Ch. 2, we find solutions of the
continuity equation for the dust flow from a point source of dust (such as a
comet or an asteroid). We also consider here the spatial distribution (in the
coordinate space) of dust particles from such a source. In Ch. 3, we
generalize those analytical solutions for evaporating dust particles, whose
radii decrease with time. Ch. 4 deals with dust flows produced by numerous
real sources (217 comets and 5000 asteroids). In Ch. 5, we summarize
our results and discuss the underlying assumptions as well as further work.

\section*{\bf 2. INTEGRALS OF MOTION}
\vskip 6pt

The rates $~v_i$ (i.e. ${da /dt}, {de /dt},...$) for
the PR-drag have been given in a number of papers
(\eg, Burns et al. 1979; Liou, Zook \& Jackson 1995):
$$
{{d a \over d t}}= \alpha{{2+3e^2\over a\left(1-e^2\right)^{3/2}}},
\eqno (4)%
$$
$$
{{d e \over d t}}= \alpha{{5e\over 2a^2\left(1-e^2\right)^{1/2}}},
\eqno (5)%
$$
\noindent
where $\alpha = - \beta{{GM\over c}}$;
$M$ is the mass of the Sun, and $\beta$ is the ratio of the solar
radiation pressure force to the gravitation force:
$$ {\beta={ 3\,L\,Q\over 16\pi G\,M\,c\,\rho\, l}}
\eqno(6)$$
where ${L\,}$ is the solar luminosity, ${G\,}$ is the gravitational constant,
${c\,}$ is the speed of light, and ${l\,}$,
${\rho\,}$ are the particle's radius and density, accordingly. The
coefficient ${Q\,}$ depends on the optical properties of the particle
and the  ratio of its size to the wavelength of light; for a perfect particle
${Q=1\,}.$ For an asteroidal dust particle, we adopt $2l=12$ $\mu$m and 
$\rho=2.7~{\rm g/cm^3}$, so that $\beta = 0.037$ (e.g., Dermott et
al. 1994).

Along the particle's trajectory, one gets
$$
{{d e \over d a}}= {{5e\left(1-e^2\right)\over 2a\left(2+3e^2\right)}}.
\eqno (7)%
$$
By integrating this equation, the particle's
trajectory in $a,e$-space is easily obtained to be
$$
{{a(1-e^2)\over e^{4/5}}}={\rm const}\equiv {C_1},  \eqno (8)%
$$
in agreement with Wyatt \& Whipple (1950). This integral of motion implies
a continuity of the particle's trajectory in ($a,e$)-space. The trajectory
is shown in Fig. 1.

{\it The flow of particles} along evolutionary trajectories shown in Fig.1
is governed by an equation (derived in the Appendix) that depends  only on
the variables $a$ and $e$ ($i={\rm const}$):
$$
f_1(a,e){{\prt n\over\prt a}} + f_2(e){{\prt n\over\prt e}} =
f_3(n,e), \eqno  (9)
$$
where $f_1=2a\left( 2+3e^2\right) $, $f_2=5e\left( 1-e^2\right) $, and $%
f_3=n\left( 6e^2-1\right)$. 
The value of $f_3$ is given here for the case when  the particle size does 
not depend on the orbital parameters $a$ and $e$ (for a more general case, 
 see Appendix).
The characteristics of the partial differential equation (9) (the integrals
of motion) are $C_1$ found above [Eq.(8)] and $C_2$ given by
$$
C_2=ne^{1/5}\sqrt{1-e^2}.\eqno  (10)
$$
The number density $n$ of dust particles along the particle trajectory in
$a,e$-space is shown as a function of semimajor axis $a$ in Fig. 2
for various $C_2$. The physical meaning of the integrals
$C_1$ and $C_2$ is the {\it conservation of the particle's flux} along
the flow under the P-R effect (note that the flux of the particles is
conserved in conditions when the energy is not).
$C_1$ and $C_2$ as the integrals of motion for the particle's flow under
the P-R drag qualitatively play the same role as the Tisserand criterion does
in the process of gravitational scattering (see GOM). As is shown in GOM,
gravitational scattering
results in ``jumps" in the particle's coordinates in $a,e$-space. However,
those jumps obey the Tisserand criterion implying the conservation of the
particle's energy in the planet's coordinate system 
under the gravitational scattering, which is an elastic
process. For a particle moving under the P-R drag alone, the energy is {\it
not} conserved and the Tisserand criterion obviously breaks down,
but the role of the integrals of motion is taken by $C_1$ and
$C_2$. Meanwhile for the particles which experience, along with the P-R drift, 
the (in fact, instantaneous) gravitational scattering by a planet,
the Tisserand criterion holds as well. Indeed, during a short time interval of
 a gravitational scattering,  the PR-drag practically does not 
change the particle's energy. 

The integrals $C_1$ (8) and $C_2$ (10) fully describe a stationary flow
of dust particles produced by an arbitrary source of dust. As for the dust
distribution in the coordinate space, it can be transformed from that in
$(a,e)$-space with the help of $C_1$ and $C_2$ by using the Haug (1958)
double integral. This numerical procedure is rather complicated but, if
we constrain ourselves to {\it point sources} in $(a,e)$-space
(such as comets and asteroids with known $a$ and $e$), the dust flows
from such sources can be analyzed by means of a much more simple
analytical technique to be presented in the next Section.

\section*{\bf 3. PARTICLE NUMBER DENSITY DISTRIBUTION\\ (Analytical Solutions
for a Point Source)}
\vskip 6pt

For a point source in $(a,e)$-space, two simple, 1-D continuity equations 
apply:
$${\partial\over\partial a}\Bigl(n(a)\cdot v_a\Bigr)=0,\eqno(11)$$
$${\partial\over\partial e}\Bigl(n(e)\cdot v_e\Bigr) =0.\eqno(12)$$
In orbital coordinates, 1-D distributions of the particle number density 
along the $a$-axis and $e$-axis are given, in the stationary state,
accordingly by
$$ n(a)=C {{a\left(1-e^2\right)^{3/2}\over\alpha\left(2+3e^2\right)}},
\eqno (13)
$$
$$
n(e)={C}{C_1^2}{{2e^{3/5}\over5\alpha\left(1-e^2\right)^{3/2}}}.
\eqno (14)
$$
Here $C$ is a constant to be found from boundary conditions, and $e$ in
equation for $n(a)$ can be eliminated with the help of the integral of motion
$C_1$ that describes, for a point source, the trajectory of motion of the
entire dust flow.

A transform from the space of orbital elements $(a,e)$ to $r$-space can
be done using known analytical expressions (Sykes 1990).
In $r$-space, the particles having orbital elements $\{a,e,i\}$, form
a rather sophisticated torus-like cloud, whose number density is described by
(Sykes 1990):

$${ P(r,\phi)=S(r)\cdot\Theta(\phi)},\eqno(15) $$
where
$$ S(r)={C_s\over{a^2}r}{1\over\sqrt{{e^2}-{({r\over a}-1)}^2 }},
\eqno(16) $$
$${ \Theta(\phi)={1\over 2{\pi^2}\sqrt{{\cos^2}\phi
-{\cos^2}i}}} \eqno(17) $$
with the limits  $\ { a(1-e)\le r \le a(1+e)} \ $ and
$\ { -i \le \phi \le i} \ $.
For a volume density near the ecliptic plane,
 $\ \it{ \phi=0}\ $ and $\ { P(r,\phi)\propto
(\sin\,i)^{-1}} \ $. It is convenient to introduce
the surface density
$${ \sigma(r)=\int \limits_{-h_0}^{h_0}P(r,\phi)dh}, \eqno(18)$$
where $\ { dh=r\, d\phi} \ $ and therefore

$${ \sigma(r)=C_\sigma\cdot S(r)\cdot r} \eqno(19) $$
with
$$ {C_\sigma=\int \limits_{-i}^{i}\Theta(\phi)d\phi}.\eqno(20)$$
It is worth mentioning that functions (16) and (18) diverge at the
pericenter and apocenter points. The total mass of dust
particles having the same orbital parameters $(a,e,i)$ in a thin
spherical region between the radii $(r_1, r_2)$ is given by:
$$ {\Delta\,m =\pi(r_1+r_2)\int \limits_{r_1}^{r_2}
\sigma(r)\,dr }. \eqno(21)$$
As a result of integration, one gets:
$${\Delta\,m(a,e) =C_{\sigma}{\pi(r_1+r_2)\over a}
\Bigl [\arcsin{1\over e}({ r_2\over a}-1)
-\arcsin{1\over e}({ r_1\over a}-1)\Bigr ] } \eqno(22)$$
where $\ { a_2 (1-e_2) \le r_2\,;\,a_1 (1+e_1) \ge r_1 } \ $.
If the region under consideration, $(r_1,r_2)$, includes the pericenter or
apocenter of the  dust particle orbit, the integration should be done as
far as the pericenter or apocenter, accordingly.
Equation (22), as opposed to eqs. (16) or (18), does not involve
divergences and is much more convenient in use.   

The distribution of the integrated surface density in the dust particle torus
with different $(a,e)$ is shown on Fig. 3.
For a dust plume produced by any given source of dust, one can find, by
using the solution (14) for $n(e)$ of the continuity equation (12),
the total dust mass in any region $(r_1,r_2)$ of the plume (taking
into account the comment on the integration limits) is:
$$ \Delta\,M = \int \limits_{e_1}^{e_2}
n(e)\cdot\Delta m(a,e)\,de . \eqno(23)$$
The result is given by
$$ \Delta\,M =C_m\cdot\pi (r_1+r_2) \int \limits_{e_1}^{e_2}
{1\over \alpha e^{1/5}\sqrt{1-e^2}} $$
$${\times\biggl\{\arcsin{1\over e}
\Bigl [{ r_2(1-e^2)\over C_1 e^{4/5}}-1 \Bigr]
-\arcsin{1\over e}\Bigl [{ r_1(1-e^2)\over C_1 e^{4/5}}-1 \Bigr]
\biggr\}\,de}.\eqno(24)$$
The integration limits are easily found by comparing the location of
the apocenter or pericenter of particle orbit relative to the boundaries
of the region under consideration, $(r_1,r_2)$. Thus the problem of
a transform from the space of orbital
elements to the $r$-space is solved by reducing it to a comparatively simple
single integration to be done numerically.

The above technique can be applied to both stationary and
non-stationary flows. The source of dust can appear owing to an
instantaneous ejection, e.g. be associated with a collision of asteroids
(Sykes 1990; Mann et al. 1996). The resulting drift
of dust particles (assumed to have the same size and located initially
as a point source at $a_0,e_0$) is shown, for three
instants of time, in Fig. 3.

Let us address a situation when the source of dust has recently
appeared  and continues to operate. This happens, for example, when 
a new comet appears in a given point of $(a,e)$-space as a result
of a `jump' due to the strong gravitational scattering of that 
comet by a planet. 
Suppose that the first dust particles ejected from such a source have reached
the semimajor axis $a_{min}$.
In order to derive the density distribution in the formed dust cloud,
it is sufficient to integrate equations(23) and (24) down to $e_{min}$, and
not through the entire dust plume.
The value of $e_{min}$ can be found from eq. (8) at the known $a_{min}$.
The appearance of this dust plume is shown, as a function of time, 
in Fig. 4. Fig. 5 demonstrates a
steady-state plume formed from various sources of dust particles. As can
be seen, small eccentricities result in the surface density $\sigma(r)\approx$
const, which implies that the {\it volume density} (i.e. accounting for
inclination $i$) is $n(r) \propto r^{-1}$. This result reproduces the 
well-known
characteristic density run of the interplanetary dust. If the source of dust
has an eccentric orbit (such as the Encke comet), the density run is much
steeper. The boundary between the plateau and cut-off of all three curves
on Fig. 5 is located at the pericenter of the source of dust,
$q=a(1-e)$, which
is due to the fact that the pericenter of a particle experiencing the P-R
drag undergoes the least shift compared to its semimajor axis and apocenter.

\section*{\bf 4. INTEGRALS OF MOTION FOR EVAPORATING PARTICLES}
\vskip 6pt
Here we deal with a situation, important in the dynamics
of the interplanetary particles, when one needs to take into account an
evaporative effect of the solar wind upon a particle (see e.g. Johnson 1991).
We describe the mass loss of a particle of mass $m$  by
$$
{d m \over d t}=- m_s \sum {N_i Y_i}.        \eqno (25)
$$
where $m_s=\mu_s {m_p}$ is the mass of molecules sputtered from the dust
particle's surface, $\mu_s$  is the molecular weight, $m_p$ is the proton
mass, $N_i$ is the number of particles of the $i$-th kind in the solar wind
colliding with the dust particle, and $Y_i$ is the sputtering coefficient,
i.e. the number of moleculs sputtered from the dust particle by one wind
particle. If the dust particles that drift  to the  Sun owing to the PR-drag
have small eccentricities, the particle's size, $l$, changes according to
a simple  law
${l \propto a^{\lambda}}$, where
the power law index $\lambda$ is given by (Shestakova 1994):
$$ {
\lambda={J_0\over 6(1+J_0)}{\mu_s\over \mu_w} Y_s}. \eqno (26)
$$
Here $\mu_w$ is an average molecular weight of the solar wind,
$ J_0 = \dot M c^2/L=0.3$ is the ratio of the
corpuscular luminosity of the Sun to the radiative one $L$, and
it is assumed that $ Q=1$. If one only accounts for the contribution
of protons and $\alpha$-particles to the dust particle's mass changes,
one finds (Shestakova 1994):
$$
      \lambda=0.032\mu_s Y_s, \eqno (27)
$$
which yields $\lambda= 0.025$ for silicates, and $\lambda= 0.56$ for ice
(H$_2$O). Bearing in mind that both $\beta$ and $\alpha \propto 1/l\propto
a^{-\lambda}$,
equation (9) takes the form (see Appendix):
$$
{2a(2+3e^2){\partial n\over\partial a}+5e(1-e^2)
{\partial n\over\partial e}=n\Bigl[6e^2(1+\lambda)+\lambda-1
\Bigr]}.\eqno(28)
$$
From this equation, the first integral given by eq. (8) is left without any
changes, which implies that the dust particle subjected to evaporation
continues to move along the same $a,e$-trajectory. Meanwhile the second
integral given by eq. (10) is modified into the following equation
(see the Appendix):
$$
C_2=ne^{1/5-(4/5)\lambda}(1-e^2)^{\lambda+1/2}.  \eqno (29)
$$
Such a modification results from an increase of the particle's
drift velocity
in the $(a,e)$-space owing to a decrease of the particle's size due to
evaporation.

The effect of the particle's evaporation can also be accounted for while 
deriving the densities $n(a)$ and $n(e)$ in eqs. (13) and (14).
In order to determine the mass of dust in any region $(r_1, r_2)$, one needs
to account for the dependence of $\alpha$ upon the function $l(a,e)$
in the integrand of eq. (24).
Besides, one needs to account for a decrease in mass of the evaporating dust
particle. As a result, a factor $l^3$ (or $a^{3\lambda}$),
with a corresponding change in the constant $C_m$, appears in the 
integrand of eq. (24). Fig. 6 demonstrates a `plume' formed from
a typical asteroid for different $\lambda=0; 0.05; 0.125; 0.25$.

\section*{\bf 5. SOURCES THAT FEED THE ZODIACAL CLOUD}
\vskip 6pt
As the sources of dust, we have incorporated the first 5000 numbered
asteroids (ITA 1994), which comprise several
distinct families (Zapalla et al. 1995) as well as
217 comets with $e<1$ taken from
Marsden \& Williams (1992). The positions of these sources on the
diagram $(\sin i, q)$ are presented on Figs. 7 (asteroids) and 8 (comets).
One can see that the asteroid pericenters tend to
concentrate in the two regions  1.7-2.15 AU and 2.6-2.9 AU.
The volume density $n(r)=\sigma/(r\sin i)$, where $r\sin i$ is the thickness
of the dust plume.  Therefore the lower the position of an asteroidal
group in Fig. 7 (small $\sin i$), the larger is its contribution to
$n(r)$ in the ecliptic plane. As for comets, their pericenters are
distributed more or less uniformly at $0.5\lax q\lax 2.5$ AU.

In order to compute the contributions of asteroidal and comet sources
to the volume density of the interplanetary dust, we adopt the simplest
assumptions: (i) all 5000 asteroids produce dust particles with the same
rate and (ii) comets produce dust particles whenever they pass pericenter
so that their contribution into the zodiacal cloud
is $\propto 1/(T q^\gamma)$, where $T$ is the comet orbital period
and typically $\gamma=2$. We neglect the
comets (sungrazers) with $a(1-e)<0.01$, assuming that the dust produced
so close to the Sun evaporates very soon.

To visualize easily any deviations of $n(r)$ from the
`standard' solution $n(r)=r^{-1}$ while making comparison with the data
on the volume density distribution of interplanetary/zodiacal dust, it is
convenient to deal with $n(r)r=\sigma/\sin i$.
The expected contribution of the above 5000 asteroidal sources to $n(r)r$
is shown in Fig. 9. It can be seen that the cut-off at $r\simeq 1.7-2.0$ AU,
associated with the asteroidal distribution in $q$, is mainly due
to the asteroids belonging to the inner and middle groups.
We have also computed $n(r)$ due to the 1000 largest asteroids with a result
that differs insignificantly from the entire sample. Therefore, our
neglecting the asteroidal distribution in sizes is not critical for
our conclusions.
The expected contribution of the above 217 comet sources to
$n(r)r$ is shown in Fig. 10. It can be seen that the density run
does not have kinks at $r>0.8$ AU and is much
steeper compared to what is due to the asteroidal sources (see Fig. 9).

Let us consider the combined cometary and asteroidal contributions
into the interplanetary dust density. The total value of $n(r) r$ for
a mixture of 25\% (in number density) cometary and 75\% asteroidal components
is shown in Fig. 11.

The density run has been approximated by $n(r) \propto r^{\delta}$.
The power law indices $\delta$ at different $r$ are given in Table 1
for a particular $\gamma=2$.

\centerline{\%\% EDITOR: Put Table 1 here \%\%}

     We have found that the results for $\gamma=1$
and  $\gamma=3$ are qualitatively similar:
the ratio of the volume densities (asteroids/comets)
has been found to be (65\%/35\%) at
$\gamma=1$ and (80\%/20\%) at $\gamma=3$. Accounting for evaporation of
dust particles changes the results in a more substantial way:
if one takes $\lambda_a$=0.025 for asteroidal dust and
$\lambda_c$=0.56 for cometary dust (Shestakova 1994) then
one gets the power law index $\delta=-1.3$  in the region $1.0\le r 
\le1.7$ AU
when the ratio of the volume densities (asteroids/comets) equals 45\%/55\%.

The asteroidal/cometary  mass ratios, which provide
the observed $\delta=-1.3$ for different possible
values of $\gamma$, $\lambda_a$, and $\lambda_c$, are given in Table 2.

\centerline{\%\% EDITOR: Put Table 2 here \%\%}

The mass ratios presented in Table 2 for spherical layers at various $r$
 differ from the ratio of volume densities in the ecliptic plane
(the 4th line of Table) because of different inclinations of the
cometary and asteroidal sources of dust and, therefore, of different
thickness  of the dust cloud.
The last column of Table 2  is fairly consistent with the estimations
for asteroidal/cometary ratios obtained from the IRAS observations
while modelling the shape of the zodiacal cloud
(Liou, Dermott, \& Xu 1995). It is worth noting that
the evaporative power law index $\lambda_a$ influences those results
insignificantly: if one puts  $\lambda_a$=0.0, the data in the last column
of Table 2 would change by no more than 1\%.

\section*{\bf 6. DISCUSSION AND CONCLUSIONS}
\vskip 6pt
The continuity equation written in the space of orbital elements, which we
have proposed to use as a tool for studying the dynamical evolution of
interplanetary particles, is very effective not only for gravitational
scattering (Gor'kavyi, Ozernoy \& Mather 1996,
GOM) but, as is shown above, for the chief factor of dynamical
evolution of the interplanetary dust cloud -- the Poynting-Robertson effect.
By solving the continuity equation, we have found two integrals of motion
governed by the P-R drag and, as a result, have derived {\it the density
run for a quasi-stationary state of the interplanetary dust cloud.}

Based on our approach, we are able to compute the formation of a dust
cloud  from many cometary, asteroidal, etc. dust sources, either observed or
assumed to exist, by adopting
various reasonable assumptions about the rates of dust production.
As an outcome, we get the theoretical model for dust distribution
both in radius and latitude, in the Solar system.
We call `the reference model' the model that only accounts for the solar
gravitational field and the P-R effect . By accounting for the other
dynamical factors, we can get corrections to the reference model.

Computations performed in the framework of the reference model yield
radial density runs of the interplanetary dust obtained under
various model assumptions and described above. They enable us to
derive some general conclusions as follows:

\begin{itemize}
\item{} The asteroidal component of the interplanetary dust
is typically characterized by
$n(r) \propto r^{-1}$ up to 1.7 AU with a sharp cut-off at $r\gax 2$ AU
[e.g  $n(r) \propto r^{-7}$ at $r\sim 3$ AU]. The predicted cut-off is
entirely consistent with what is observed for the zodiacal cloud
and the main core-population of the dust in the Solar system (Divine 1993);

\item{} The power law index $\delta$ in the distribution of the volume
density of the zodiacal dust particles, $n(r)\propto r^{\delta}$, depends
on several factors:
\begin{itemize}
\item{} at radii smaller than the pericenter of the source of dust,
($r<a(1-e)$), the index $\delta\equiv \delta_1$ differs significantly
from that, $\delta\equiv \delta_2$, at radii larger than the pericenter
of the source of dust, ($r>a(1-e)$);

\item{} both $\delta_1$ and $\delta_2$ depend strongly on eccentricity
of the source: as
 $e$ increases from 0 to 1, $\delta_1$ increases from $-1$ to 0 whereas
the value of $\delta_2$ increases from $-\infty$ to $-2.5$;

\item{} accounting for the particle's evaporation,
both  $\delta_1$ and $\delta_2$ tend to increase (see Fig. 6);

\item{} for a system of dust sources distributed in $a$ and/or $e$,
the index $\delta$ differs substantially inside and outside
the system (see Fig. 12). For such distributed sources, there is a smooth
transition between (and not a jump from)  $\delta_1$ and $\delta_2$
near the pericenter.
\end{itemize}

\item{}  In the region $r< 1.7$ AU, observations yield $n(r) \propto
r^{-1.3}$ (Divine 1993) or  $r^{-1.25}$ (Levasseur-Regourd 1996).
The observed deviation from the classical law $r^{-1}$  may be related
to the contribution of the cometary dust component. This component (which
alone would result in $n(r)\propto r^{-2.4}$) is expected to contribute
as much as
about 25\% in number density (or about 50\% in mass) to the dust cloud
near Earth;

\item{} The ratio of the cometary dust contribution to that of the
asteroidal dust into $n(r)$ has a minimum at $r\simeq$ 1.8-2 AU and it is
large in the inner region of the Solar system ($a<1$ AU) as well as in
the outer region of the main asteroid belt ($a>3$ AU).

\item{} The dust created by the Kuiper belt should fill the inner region
of the Solar system as a more or less homogeneous layer with
$\sigma(r)={\rm const}$. At $r> 3-4$ AU that dust may dominate over the
comet and asteroidal dust components.
\end{itemize}

Let us turn to discussing how  the adopted approximations influence
our conclusions. The basic approximations of our reference model
are: (i) a steady-state character of the model and
(ii) neglect of other dynamical factors except the P-R
drag.

Approximation (i) could be easily abandoned -- we are able to compute
the non-stationary processes as well.
For instance, Fig. 4 shows different stages of dust plume formation
of a source that once appeared and continues to produce dust particles.
Therefore, this approximation does not imply any restriction in our approach
or a failure of our model; it only results from an incompleteness
of our knowledge of whether the real sources of dust are stationary or not.

Approximation (ii) implies that our solutions fail to describe
such phenomena as the resonance circumsolar dust rings near Earth (Jackson
\& Zook 1989), which may contain as many as $\sim 10$\% of the particles of
the zodiacal cloud (Dermott et al. 1994; Liou, Dermott, \& Xu 1995).
In our further work, we will fully incorporate the resonance and
related effects.

Some minor approximations in our computations, such as the intensity of comets
as dust sources as a function of their distances from the Sun, and the law of
dust particle evaporation, are also worth refining. Besides, we have
ignored in the above computations the distribution of dust sources in size
and chemical abundance as well as the effects of observational selection
in asteroidal and cometary populations. However, all uncertainties in the
those efffects can be easily explored by varying the parameters of the
model for the zodiacal cloud formation. For instance, while computing
a multi-component model of the dust cloud, we need to compute the
reference model and get the structure of the dust cloud consisting of the
particles of a given size. The value of $\beta$ influences not just the
speed of a particle's motion along the  orbital trajectories
but, when $\beta$ is not too small compared to 1, one needs to account for
the change of orbital coordinates of the dust particles beginning with
the instant of their escape from the source (see, e.g., Jackson \& Zook
1992).

Despite the above approximations, our reference model has many principal
advantages. Being semi-analytical it is flexible enough to be able to
compute the dust cloud formation from a large number of various sources
having an arbitrary distribution in size, composition, etc. It takes a short
CPU time (e.g., 4 min for 217 comets and 1 hour for 5000 asteroids with a
486/DX4-100) to compute the typical reference model.
Therefore it is feasible to deal with multi-variant dust population sources
while fitting the available cometary, asteroidal, and
far infrared COBE data. Then one could employ our
basic model of the dust cloud to incorporate the other effects
such as gravitational scattering of dust particles by planets
(Gor'kavyi, Ozernoy \& Mather 1997), resonance capture of dust (Jackson
\& Zook 1989; Dermott et al. 1994), particle multiple collisions, etc.

We conclude by emphasizing that our continuity equation (1) is well suited to
the study of various large
scale properties of the interplanetary dust cloud.
Our approach might be fruitful for solving a number of related
problems, including the structure of protoplanetary and circumstellar
disks.
\newpage
\section*{ Appendix\\ Integrals of motion.}
Here we derive the integrals of motion based on eq. (11).
At first, we address the case when the particle's size does not depend
upon the orbital parameters, and thus the
coefficient ${\alpha}$ in eqs. (4) and (5)
does not depend upon the semimajor axis ${a}$ and eccentricity ${e}$.
By differentiating eqs. (7) and (8), one finds
${\partial v_a / \partial a }$ and ${\partial v_e / \partial e } $,
respectively:
$${\partial v_a\over \partial a}=\alpha {2+3e^2
\over{(1-e^2)}^{3/2}}{\partial\over\partial a }
\Biggl({1\over a}\Biggr)=
-\alpha {2+3e^2\over a^2{(1-e^2)}^{3/2}}, \eqno(A1) $$
$${\partial v_e\over \partial e}=\alpha {5
\over 2 a^2}{\partial\over\partial e }
\Biggl ({e\over (1-e^2)^{1/2}}\Biggr)=
\alpha {5\over 2 a^2{(1-e^2)}^{3/2}}. \eqno(A2) $$
By substituting eqs. (7), (8), (A1), and (A2) into eq. (11), one gets:
$${\alpha(2+3e^2)\over a{(1-e^2)}^{3/2}}{\partial n\over
\partial a} -{n\alpha(2+3e^2)\over a^2{(1-e^2)}^{3/2}}+
{5\alpha e\over 2a^2{(1-e^2)}^{1/2}}
{\partial n \over\partial e } $$
$$+{5n\alpha\over 2a^2{(1-e^2)}^{3/2}}=0.\eqno(A3) $$
Having multiplied eq. (A3) by ${2\alpha^{-1}a^2(1-e^2)^{3/2}}$,
one finds finally:
$$2a(2+3e^2){\partial n\over\partial a}+5e(1-e^2)
{\partial n\over\partial e}=n(6e^2-1) \eqno(A4)$$
This partial differential equation implies the following ordinary
differential equation :
$${{da\over 2a(2+3e^2)}={de\over 5e(1-e^2)}=
{dn\over n(6e^2-1)} }\eqno(A5) $$

Each pair of equations comprising eq. (A5) is solved
by separating the variables to give
$${\int{1\over a}da={2\over5}\int{2+3e^2\over e(1-e^2)}
de\, }\eqno(A6) $$
and
$${\int{1\over n}dn={1\over5}\int{6e^2-1\over e(1-e^2)}
de\, }.\eqno(A7) $$
The integrals can be easily calculated to give the integrals
of motion discussed in the main text:
$${C_1 = {a(1-e^2)\over e^{4/5}}} \eqno(A8) $$
and
$${ C_2 = n e^{1/5}\sqrt{1-e^2} }. \eqno(A9) $$

Let us now address the case when the particle's size, owing to
evaporation, depend on the orbital parameters
${a}$ and ${e}$. For simplicity, we assume that the particle's radius
${l}$ changes with $a$ according to a power law. Then the
coefficient ${\alpha}\propto l^{-1}\propto a^{-\lambda} $, i.e.
$${\alpha = \alpha_0 a^{-\lambda} }. \eqno(A10)$$
In this case, one gets, instead of eq. (A1):
$${{\partial v_a\over \partial a}={2+3e^2
\over{(1-e^2)}^{3/2}}{\partial\over\partial a }
\Biggl ({\alpha\over a}\Biggr)=-\alpha_0(1+\lambda)
{2+3e^2\over a^{2+\lambda}{(1-e^2)}^{3/2}}},
\eqno(A11) $$
as well as the following equation, instead of eq. (A2):
$${{\partial v_e\over \partial e}=\alpha
{5\over 2 a^2{(1-e^2)}^{3/2}}={5\alpha_0\over 2 a^{2+\lambda}
{(1-e^2)}^{3/2}}}. \eqno(A12) $$
Having substituted eqs. (A10),
(A11), (A12), (7), and (8) into eq. (11), one gets after some algebra
$${2a(2+3e^2){\partial n\over\partial a}+5e(1-e^2)
{\partial n\over\partial e}=n\Bigl[6e^2(1+\lambda)+\lambda-1
\Bigr]}.\eqno(A13)$$
Since the l.h.s. of eq. (A13) and (A4) are equal to each other, ${C_1}$
is kept the same as in eq. (A8) for a non-evaporative case.
Meanwhile for the r.h.s.
of eq. (A13) one gets, similar to eq. (A7):
$${\int{1\over n}dn={1\over5}\int{6e^2
(1+\lambda)+\lambda-1\over e(1-e^2)}de\,, }\eqno(A14) $$
which yields the second integral of motion for (evaporating) particles
of a changing radius:
$$
C_2=ne^{1/5-(4/5)\lambda}(1-e^2)^{\lambda+1/2}.  \eqno(A15)
$$
\newpage

\def\apj    {{ApJ.{\rm,}\ }}
\def\apjl   {{ApJ.~(Letters){\rm,}\ }}
\def\apjs   {{\sl Ap.~J.~Suppl.{\rm,}\ }}
\def\aa     {{A\&A{\rm,}\ }}
\def\anrev  {{Ann.~Rev.~Astr.~Ap.{\rm,}\ }}
\def\baas   {{BAAS{\rm,}\ }}
\def\mnras  {{MNRAS{\rm,}\ }}
\def\ref#1  {\noindent \hangindent=24.0pt \hangafter=1 {#1} \par}
\def\v#1  {{{#1}{\rm,}\ }}
{}
\newpage
\centerline{Table 1. {\bf Power Law Indices $\delta$}}
$$\vbox{\offinterlineskip
\halign{&\vrule#&\strut\ #\ \cr
\multispan{9}\hfil\ \hfil\cr
\noalign{\hrule}
height3pt&\omit&&\omit&&\omit&&\omit&\cr
&\hfil Region\hfil&&\hfil Asteroidal dust\hfil&&\hfil
Cometary dust\hfil&&\hfil 75\% ast. dust + 25\% com. dust\hfil&\cr
height3pt&\omit&&\omit&&\omit&&\omit&\cr
\noalign{\hrule}
height3pt&\omit&&\omit&&\omit&&\omit&\cr
&\hfil $r=0.5-1.7$ AU\hfil&&\hfil$-1.03\pm 0.02$\hfil&&\hfil $-2.17\pm 0.38$
\hfil&&\hfil $-1.29\pm 0.06$\hfil&\cr
&\hfil $r=1.0-1.7$ AU\hfil&&\hfil$-1.04\pm 0.02$
\hfil&&\hfil$-2.36\pm 0.29$\hfil&&\hfil $-1.29\pm 0.03$\hfil&\cr
&\hfil $r\approx 2.0$ AU\hfil&&\hfil$-2.6$\hfil&&\hfil$-2.8$\hfil&&\hfil$-2.6$
\hfil &\cr
&\hfil $r\approx 2.5$ AU\hfil&&\hfil$-4.0$\hfil&&\hfil$-2.9$\hfil&&\hfil$-3.8$
\hfil &\cr
&\hfil $r\approx 3.0$ AU\hfil&&\hfil$-7.4$\hfil&&\hfil$-3.1$\hfil&&\hfil$-6.4$
\hfil &\cr
height3pt&\omit&&\omit&&\omit&&\omit&\cr
\noalign{\hrule}
\cr}}$$
\vskip 12pt
\centerline{Table 2. {\bf Mass of Asteroidal Dust /
Mass of Cometary Dust (\% / \%)}}
$$\vbox{\offinterlineskip
\halign{&\vrule#&\strut\ #\ \cr
\multispan{11}\hfil\ \hfil\cr
\noalign{\hrule}
height3pt&\omit&&\omit&&\omit&&\omit&&\omit&\cr
&\hfil && \hfil ${\gamma}=1$\hfil&&\hfil ${\gamma}=2$\hfil&&
\hfil ${\gamma}=3$\hfil&&\hfil ${\gamma}=2$\hfil&\cr
height3pt&\omit&&\omit&&\omit&&\omit&&\omit&\cr
&\hfil r\,(AU)\hfil&&\hfil ${\lambda_a}=0$\hfil&&
\hfil ${\lambda_a}=0$\hfil&&\hfil ${\lambda_a}=0$\hfil&&
\hfil ${\lambda_a}=0.025$\hfil&\cr
height3pt&\omit&&\omit&&\omit&&\omit&&\omit&\cr
&\hfil && \hfil ${\lambda_c}=0$\hfil&&
\hfil ${\lambda_c}=0$\hfil&&
\hfil ${\lambda_c}=0$\hfil&&\hfil ${\lambda_c}=0.56$\hfil&\cr
height3pt&\omit&&\omit&&\omit&&\omit&&\omit&\cr
&\hfil 1\hfil&&\hfil ${n_a /n_c}=65/35$\hfil&&
\hfil ${n_a /n_c}=75/25$\hfil&&\hfil ${n_a /n_c}=80/20$\hfil&&
\hfil ${n_a /n_c}=45/55\, ^{*)}$\hfil&\cr
\noalign{\hrule}
height3pt&\omit&&\omit&&\omit&&\omit&&\omit&\cr
&\hfil $ 0.0 - 1.0 $ \hfil&&\hfil$ 39/61 $\hfil&&\hfil $ 42/58 $\hfil&&
\hfil $ 37/63 $\hfil&&\hfil $ 19/81 $\hfil&\cr
&\hfil $ 1.0 - 3.0 $ \hfil&&\hfil$ 56/44 $\hfil&&\hfil $ 69/31 $\hfil&&
\hfil $ 75/25 $\hfil&&\hfil $ 28/72 $\hfil&\cr
&\hfil $ 3.0 - 5.0 $ \hfil&&\hfil$ 21/79 $\hfil&&\hfil $ 33/67 $\hfil&&
\hfil $ 40/60 $\hfil&&\hfil $ 5/95 $\hfil&\cr
height3pt&\omit&&\omit&&\omit&&\omit&&\omit&\cr
height3pt&\omit&&\omit&&\omit&&\omit&&\omit&\cr
&\hfil $ 0.0 - 5.0 $ \hfil&&\hfil$ 46/54 $\hfil&&\hfil $ 54/46 $\hfil&&
\hfil $ 52/48 $\hfil&&\hfil $ 21/79 $\hfil&\cr
&\hfil $ 0.95 - 1.05 $\hfil&&\hfil$ 46/54 $\hfil&&\hfil $ 58/42 $\hfil&&
\hfil $ 62/38 $\hfil&&\hfil $ 25/75 $\hfil&\cr
height3pt&\omit&&\omit&&\omit&&\omit&&\omit&\cr
\noalign{\hrule}
\cr}}$$

*) asteroidal/cometary volume densities of dust at $r=1$ AU in the ecliptic
plane.
\newpage
\centerline {\bf Figure Captions}
\blankline
Fig. 1. Changes in the particle's position in the $(a,e)$-space
 due to the Poynting-Robertson drag. Initial semimajor axis is $a_0=2.5$ AU;
 initial eccentricities are $e_0= 0.15,~ 0.29,~ 0.43,~ 0.57,~
0.71,~ 0.85$, and 0.99 from bottom to top, accordingly (the corresponding
$C_1$ are 11.15, 6.16, 4.00, 2.64, 1.63, 0.79, and 0.05).
The transforms of orbits A, B, and C to the $r$-space are shown in Fig. 3.

\blankline

Fig. 2.
The number density of dust particles along the particle trajectory in
$a,e$-space, $n$, is shown as a function of semimajor axis $a$
for various initial
eccentricities $e_0= 0.15,~ 0.29,~ 0.43,~ 0.57,~ 0.71,~ 0.85$, and 0.99 (with
$n_0=1$). The upper dashed line, ${n}={C_3\over a^{1/4}}$, corresponds to
the circular case ($e_0=0$); the lower dashed line, ${n}={C_4\ a^{1/2}}$,
corresponds to the parabolic case ($e_0=1$).

\blankline

     Fig. 3. The surface density, $\sigma(r)$, integrated over small
$(r_2-r_1=0.01$ AU) regions as a function of
radius $r$ for three different points A, B, and C shown in
Fig. 1. Initial $(a_0,e_0)$-data for those points are, respectively:
(2.5, 0.29); (1.3, 0.14); and (0.6, 0.054).

\blankline

Fig. 4. Seven stages of the evolution for the surface
density of a cometary plume having $a_0=3.0$, $e_0=0.5$.
Curves labeled 1 to 7 corresponds to the instants of time by which the
semimajor axis of particles decreases by ${1\over 7}a_0$.

\blankline

Fig. 5. The surface density $\sigma$ as a function of
radius $r$ for the $a,e$-curve of three different sources of dust
having initial semimajor axis $a_0=2.22$ and eccentricities $e_0
=0.15;~ 0.5;~ 0.85$ (curves 1, 2, and 3, accordingly). The first case is
typical for the inner asteroids, and the last case is the Encke comet.

\blankline

     Fig. 6. The surface density $\sigma$ as a function of
radius $r$ for the $a,e$-curve of sources of dust
with $a_0=2.22$ and eccentricity $e_0=0.15$
at different $\lambda=0;~ 0.05;~ 0.125;~{\rm and}~ 0.25$ (curves 1, 2, 3,
and 4, accordingly).

\blankline

     Fig. 7. 5000 asteroids (taken from ITA 1994)
plotted in the [$\sin i,~ a(1-e)$]-space.
Enhanced concentrations of points indicate at least 4 to 5 different
asteroidal groups.

\blankline

     Fig. 8. Comets in the $[\sin i,~a(1-e)$]-space (Marsden \& Williams
1992). Notations:
crosses for 97 short-periodic comets with $a<20$ AU and $i<90^
{\circ}$;
squares for 67 long-periodic comets with $a>20$ AU and $i<90^
{\circ}$;
triangles for 53 retrograde comets with $i>90^{\circ}$
(all but three comets have $a<20$ AU).

\blankline

     Fig. 9. Function $n(r) r\equiv\sigma (r)/ \sin i$ as a function of
radius $r$ for dust particles of asteroidal origin.
Curves 1, 2, and 3
correspond, respectively, to 1816 asteroids  of the inner zone ($a<
2.501$ AU); to 1271  asteroids of the mid zone ($2.501<a<2.825$ AU);
and to 1913 asteroids of the outer zone
($a>2.825$ AU). Curve 4 is the integral for all 5000 asteroids. One can
see that $n(r)r$ has a cut-off at $r\simeq 1.7-2.0$ AU.
The curve consisting of squares  corresponds to the dust produced by
all 5000 asteroids with accounting for
evaporation ($\lambda=0.025$).

\blankline

    Fig. 10. Function $n(r) r\equiv\sigma (r)/ \sin i$ as a function of
radius $r$ for dust particles of cometary origin.
Curves 1, 2, and 3 correspond, respectively, to all
comets with $\gamma=1,~ 2,$ and 3.
The solid curve (the evaporative case) corresponds to the
dust produced by comets with $\gamma= 2$ and $\lambda= 0.56$.

\blankline

     Fig. 11. Functions $n(r) r$ for cometary dust
(squares) and asteroidal dust (triangles). Each function is normalized
to unity at $r=1$ AU.
Crosses stand for a mixture of cometary component (25\%) and asteroidal
component (75\%) (\%\% are taken in number density at $r=1$ AU in the
ecliptic plane).

\blankline

     Fig. 12. Functions $\delta(r)$ for
a single comet with $a=2.22$AU and $e=0.85$ (curve 1);
50 fictitious comets uniformly distributed in $e$ within the range of
$0.6<e<0.85$ (curve 2);
a sample of 217 real comets (curve 3);
a single asteroid having $a=2$AU and $e=0.15$ (curve 4);
200 fictitious asteroids uniformly distributed in $a$ within the range of
$2<a<3$ (curve 5);
and a sample of 5000 real asteroids (curve 6).

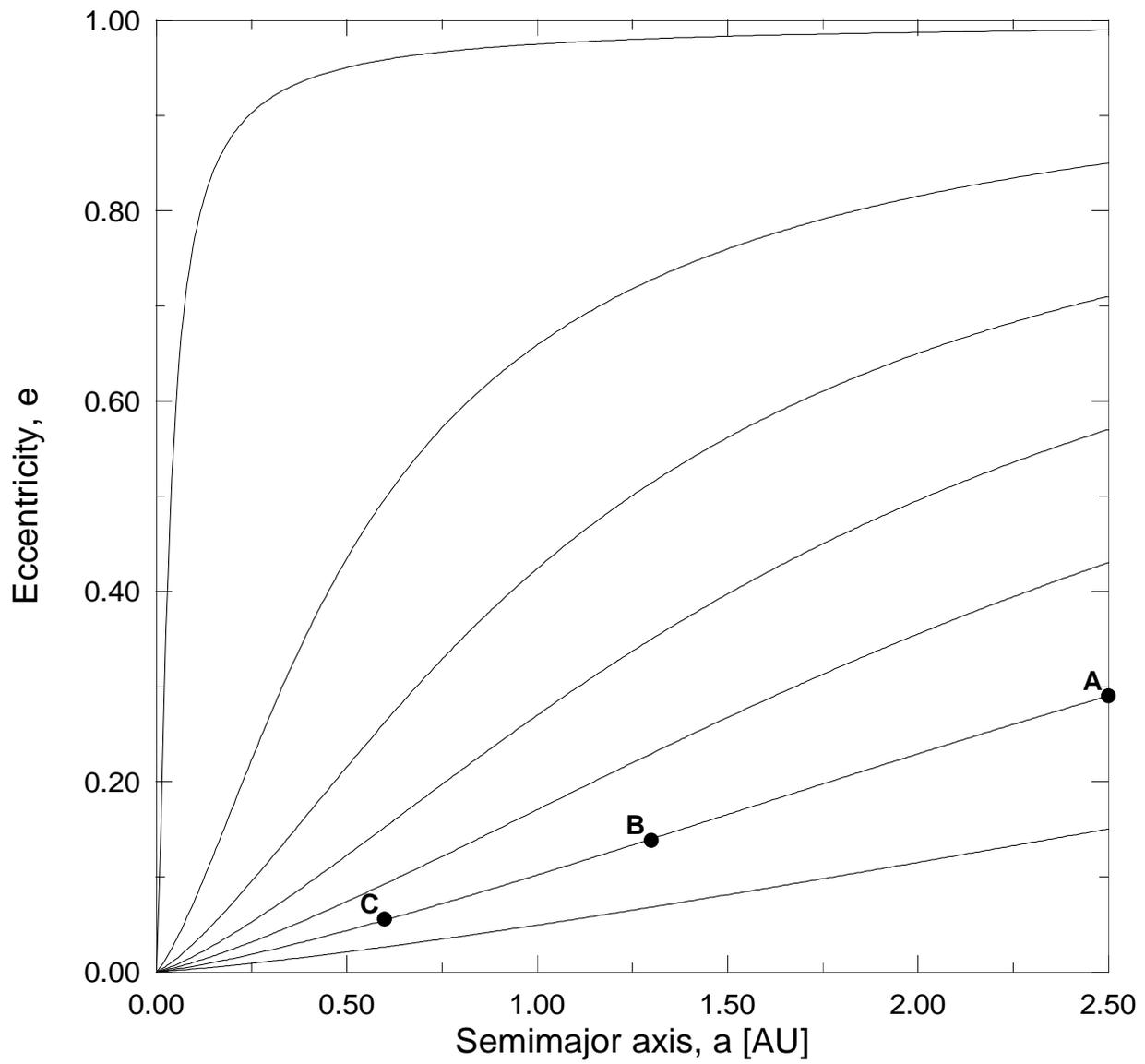
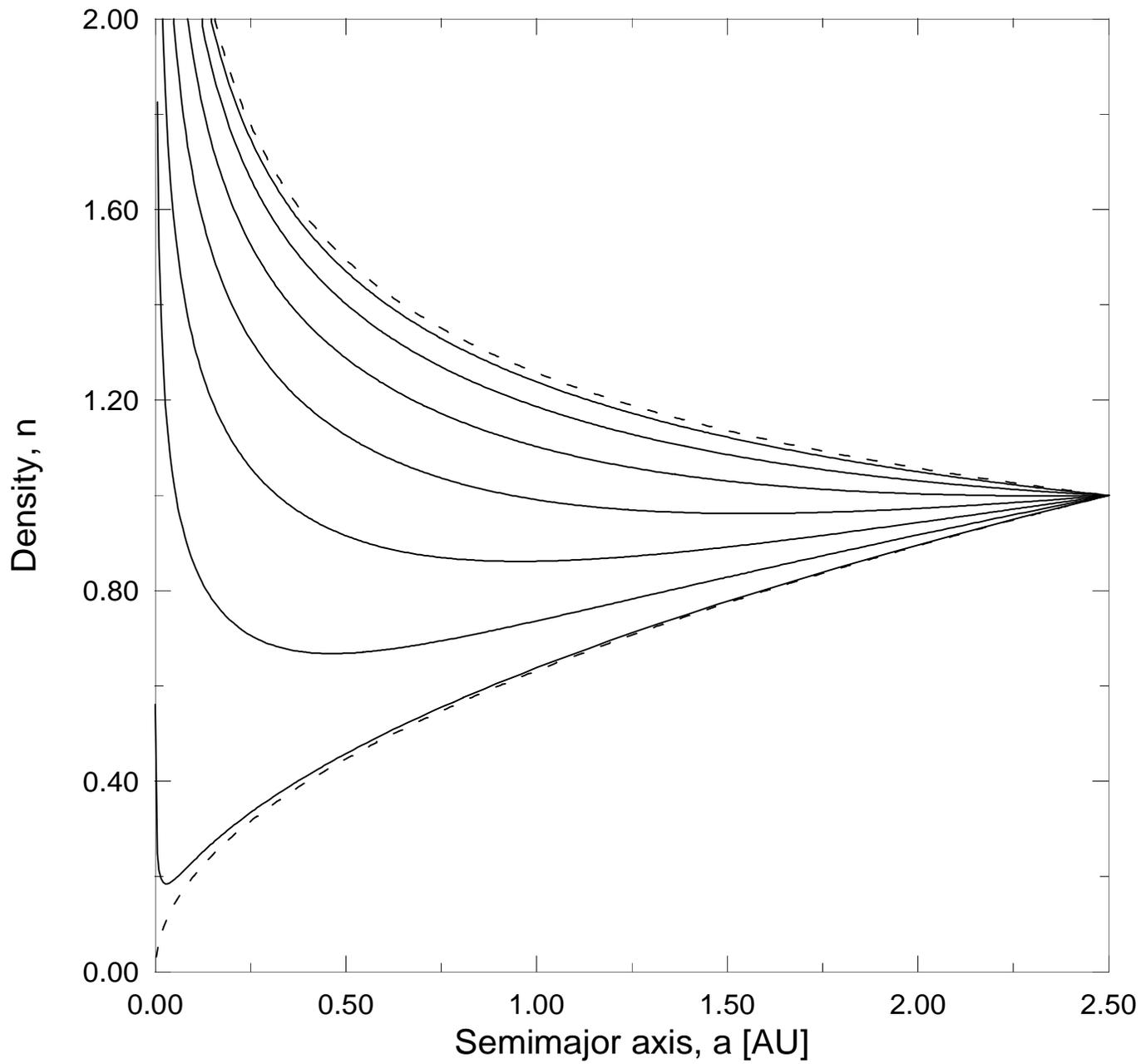
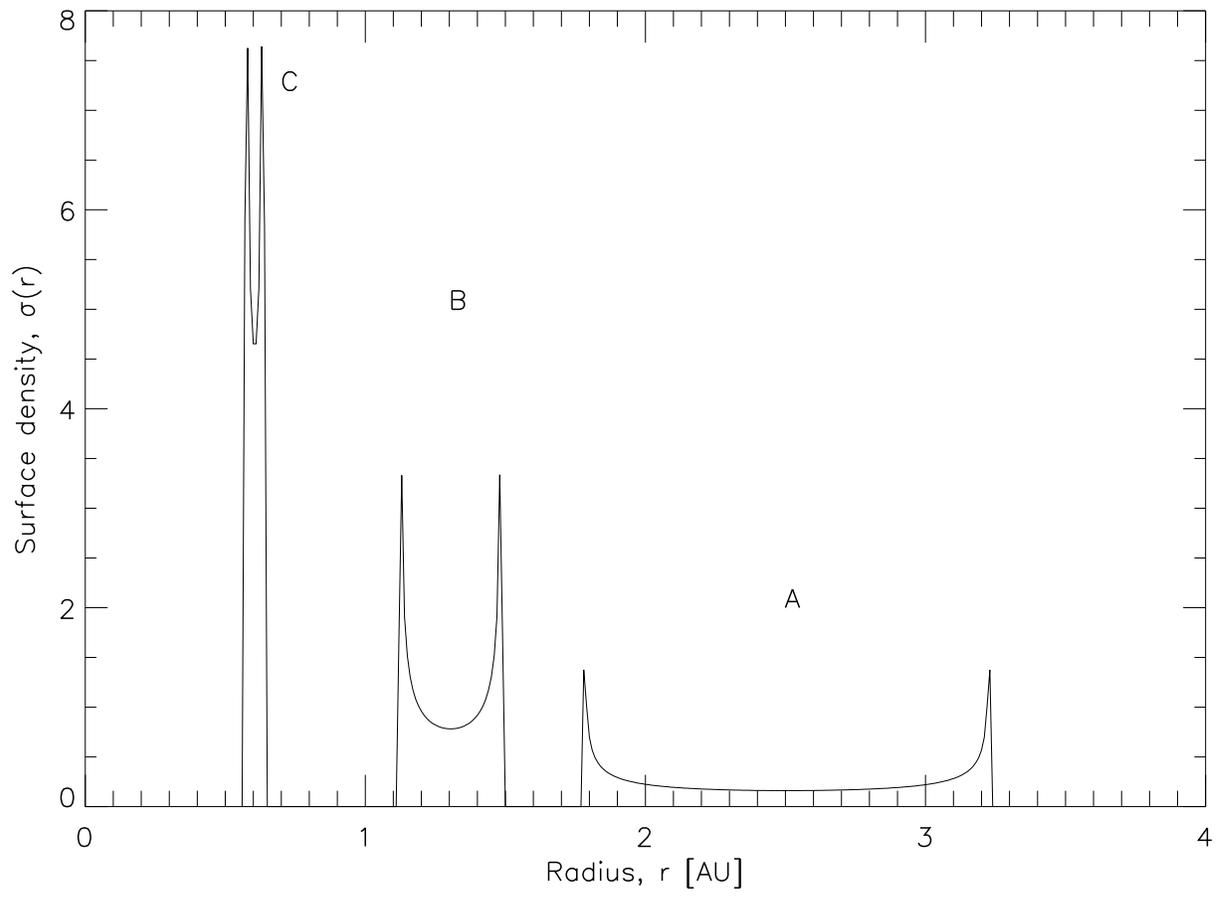
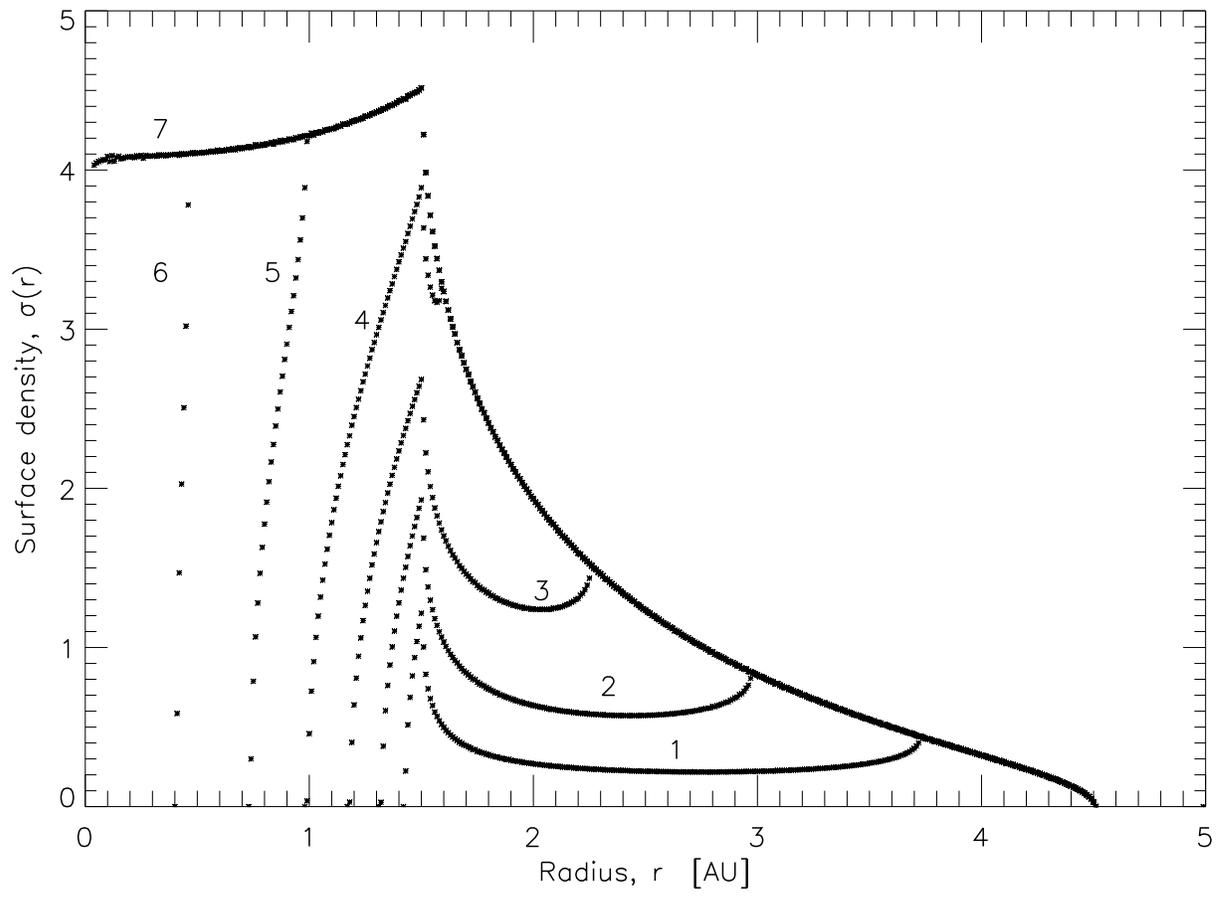
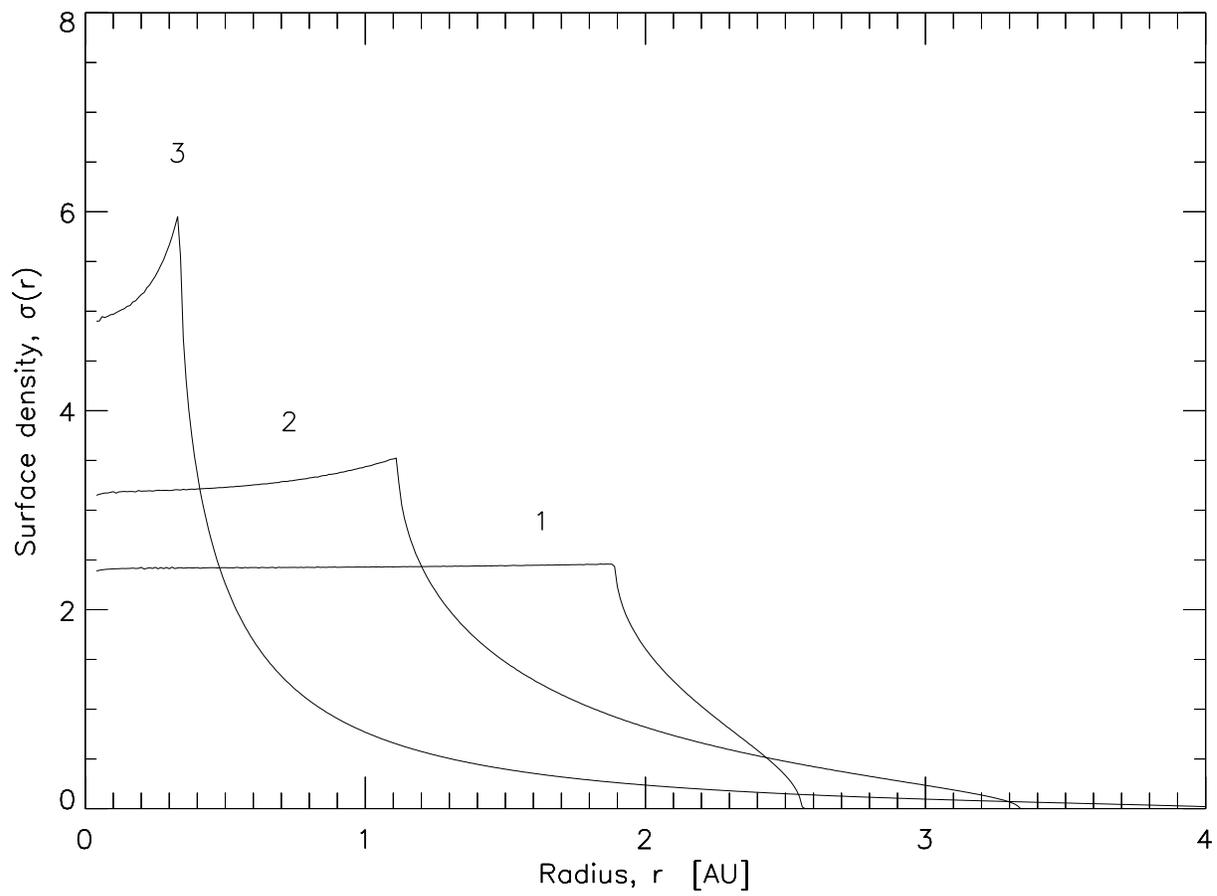
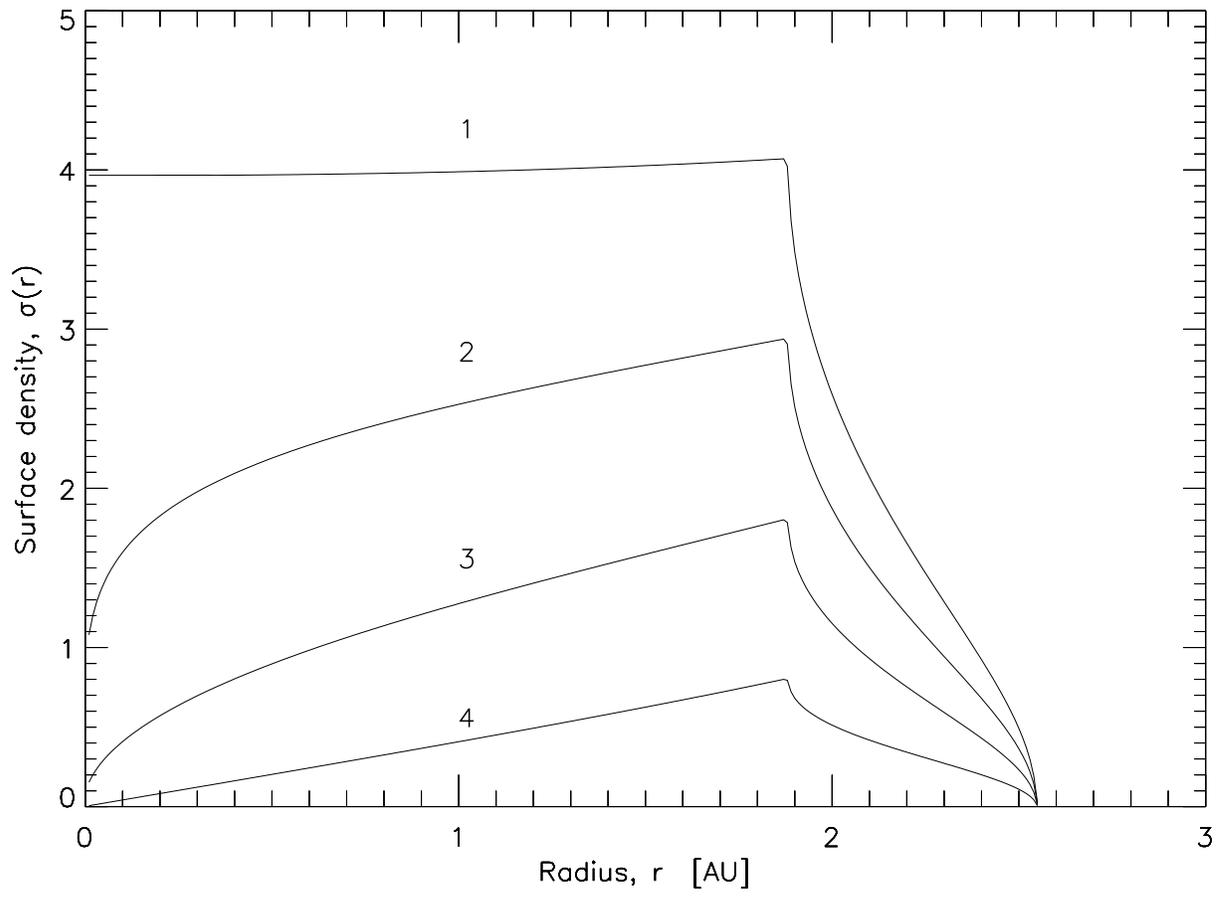
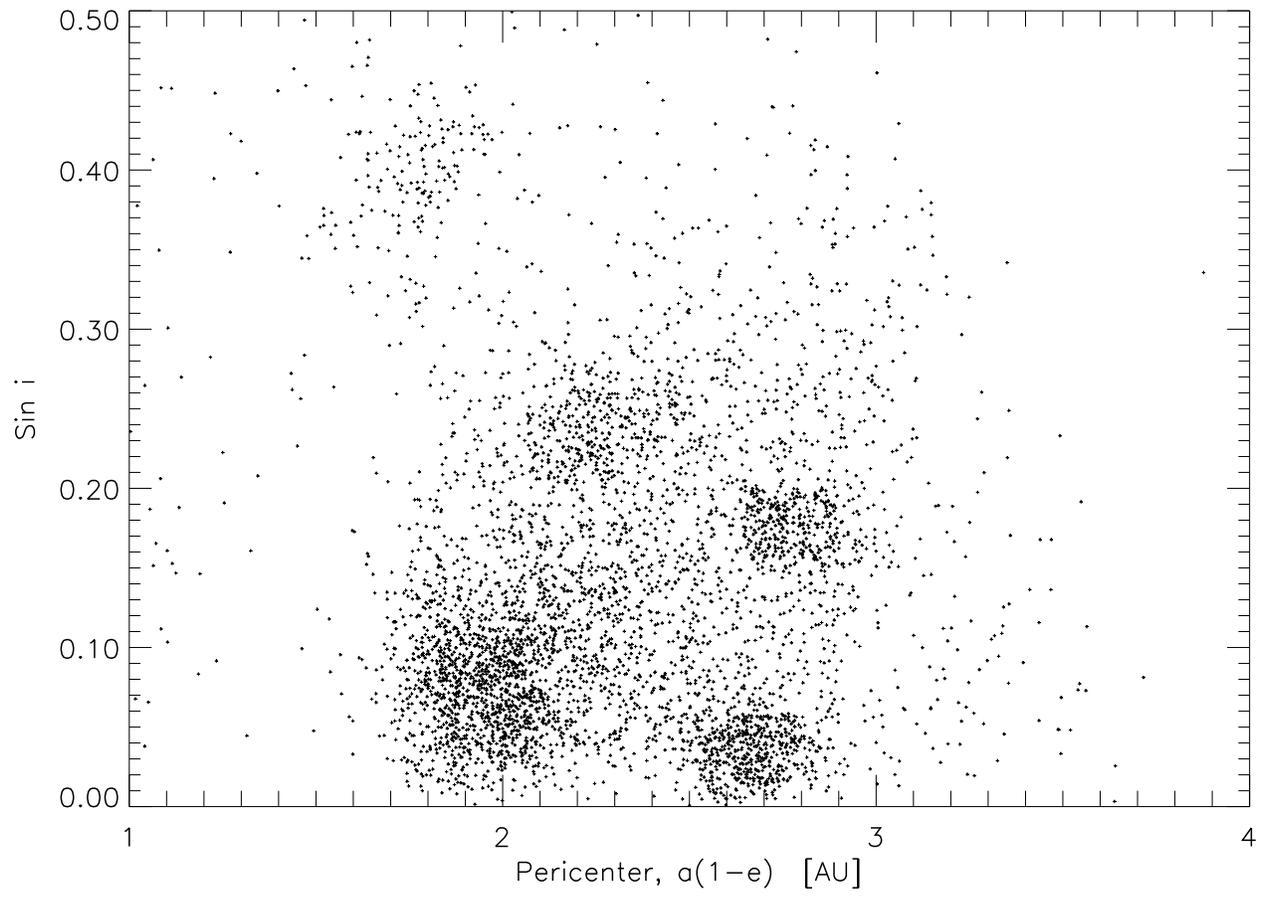
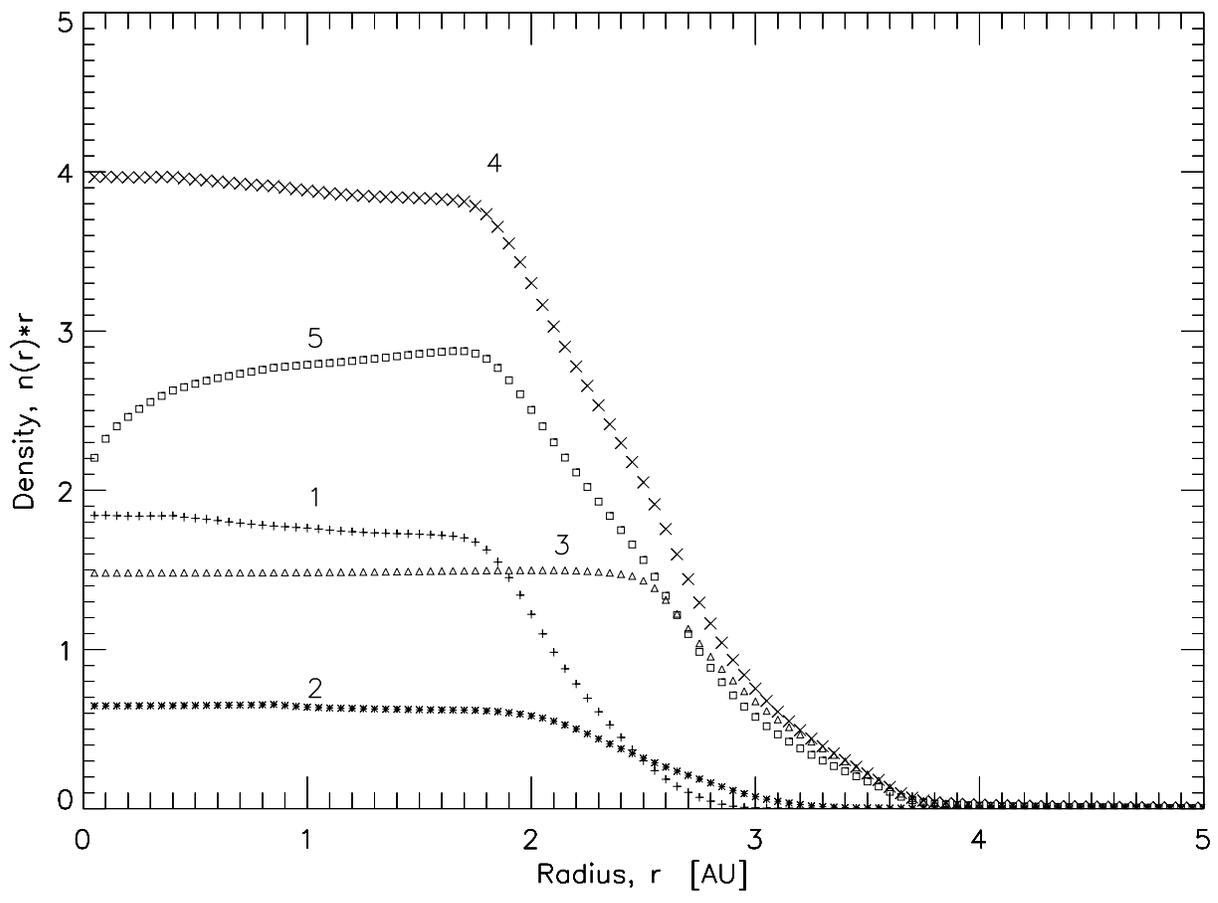
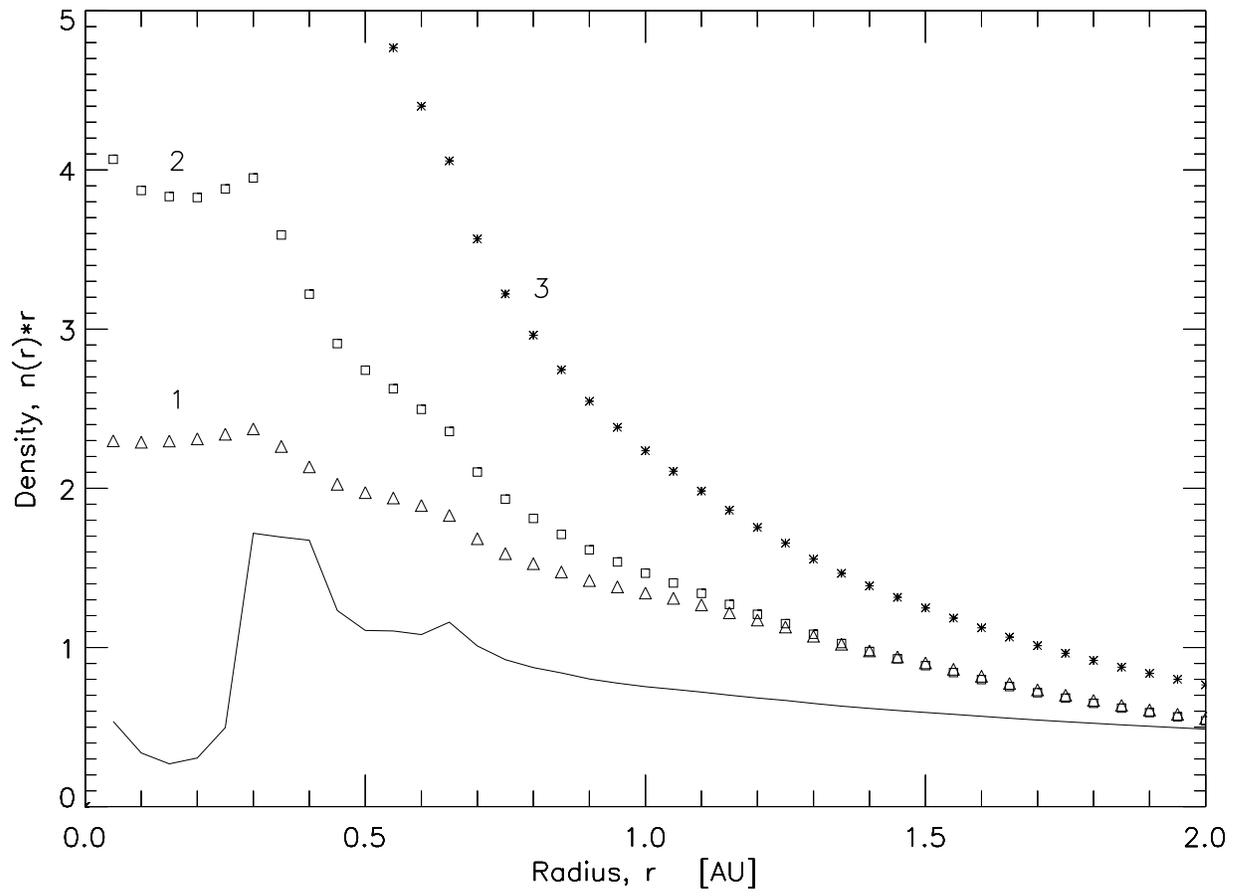
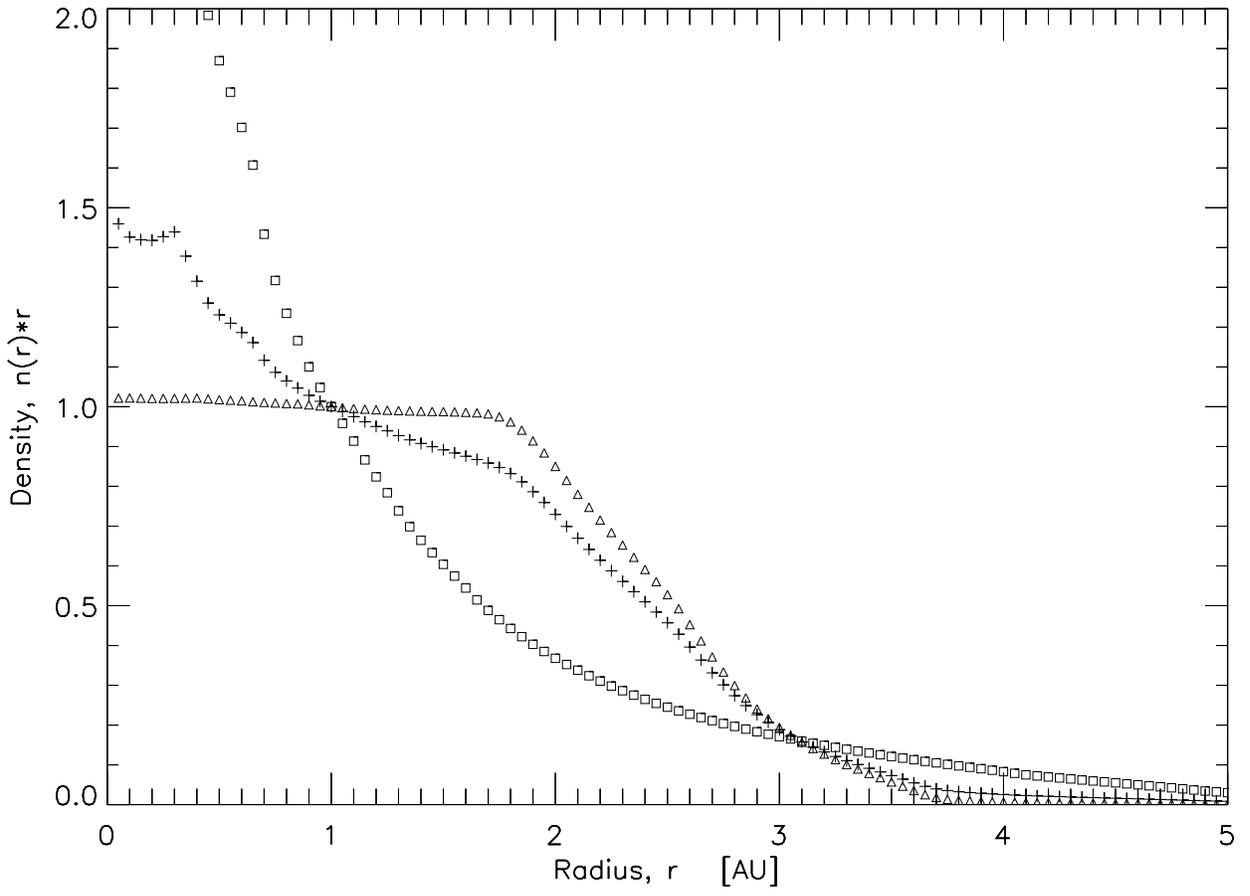
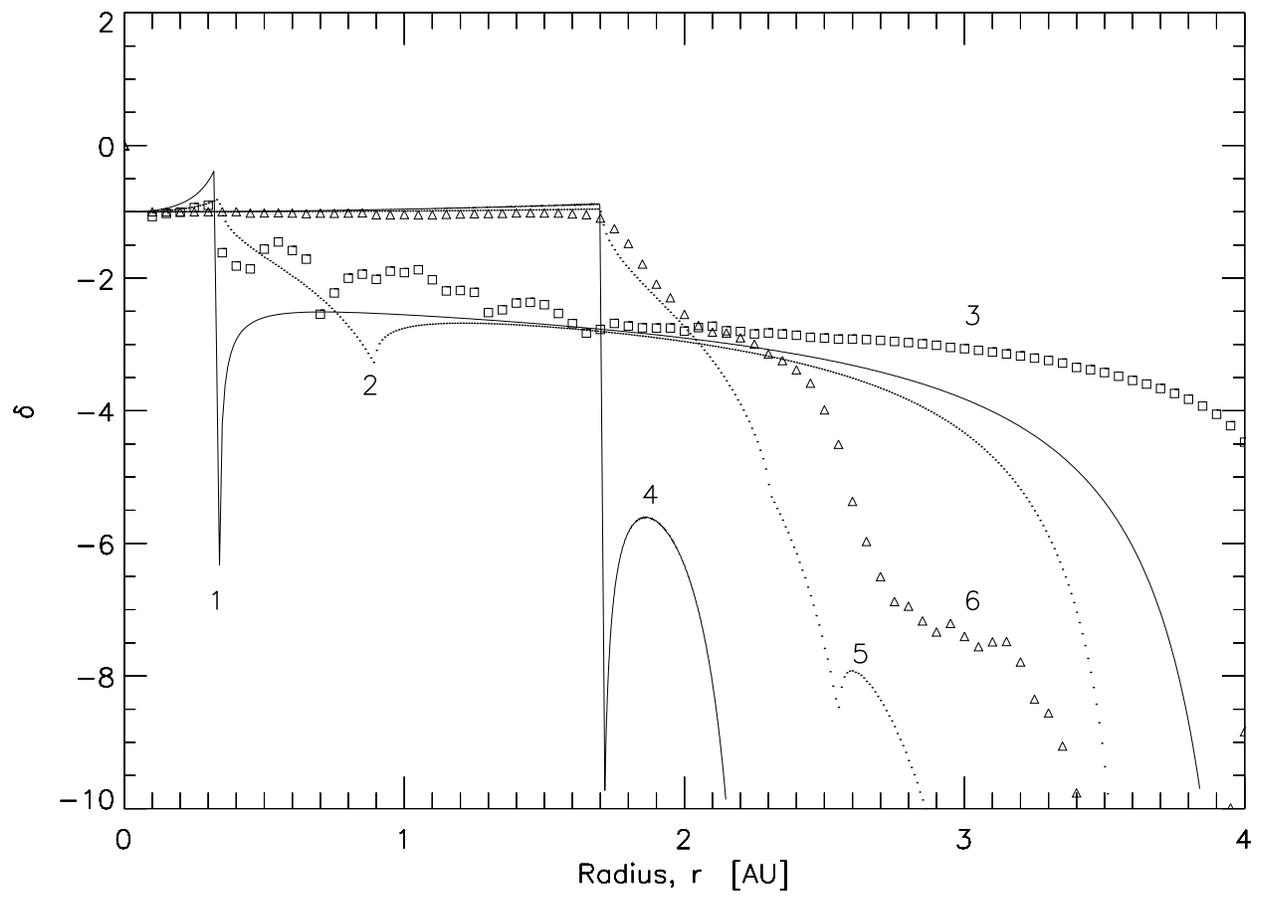
\end{document}